\documentclass[conference]{IEEEtran}

\ifCLASSOPTIONcompsoc
  \usepackage[nocompress]{cite}
\else
  \usepackage{cite}
\fi
\ifCLASSINFOpdf
  \usepackage[pdftex]{graphicx}
\else
\fi

\usepackage{amsmath}

\usepackage[ruled, linesnumbered,noend]{algorithm2e}

\usepackage{array}

\ifCLASSOPTIONcompsoc
  \usepackage[caption=false,font=footnotesize,labelfont=sf,textfont=sf]{subfig}
\else
  \usepackage[caption=false,font=footnotesize]{subfig}
\fi
\usepackage{fixltx2e}
\usepackage{float}

\usepackage[multiple]{footmisc}
\usepackage{url}
\usepackage{esvect}
\usepackage{textcomp}
\usepackage{diagbox}
\usepackage{tabularx}
\usepackage{dcolumn}
\usepackage{booktabs}
\usepackage[english]{babel}
\usepackage[dvipsnames]{xcolor}
\usepackage[utf8]{inputenc}
\usepackage[T1]{fontenc}
\usepackage{latexsym}
\usepackage{amssymb}
\usepackage{listings}
\usepackage{caption}
\usepackage{subfig}

\usepackage{longtable}
\usepackage{rotating}
\usepackage{array}
\usepackage{multicol}
\usepackage{multirow}

\newtheorem{definition}{Definition}

\makeatletter
\newcommand{\removelatexerror}{\let\@latex@error\@gobble}
\makeatother

\makeatletter
\newcommand*{\rom}[1]{\expandafter\@slowromancap\romannumeral #1@}
\makeatother

\lstdefinelanguage{code}{
  morekeywords={let,in,def,aspect,before,after,pointcut,public,privileged,protected,declare,parents, call,target,implements,throw,new,for,class,forAll,exists,Boolean,return,break,executeCaller,true,
  false,pre,if,then,else,endif,String,join,select,from,where,with,create,temporary,table,and,
  or,as,on,case,when,end,union,iterate,intersection,symmetricDifference,self,includes,function, Text,Set,boolean, show,insert, into,div,mod,Integer, by, having, on, not, var, while, continue,switch,execute,abort,delete,eval,this,input, output,process},
  basicstyle=\footnotesize\usefont{T1}{pcr}{m}{n}\selectfont,
  keywordstyle=\footnotesize\usefont{T1}{pcr}{b}{n}\selectfont,
  identifierstyle=\footnotesize\usefont{T1}{pcr}{m}{n}\selectfont,
  commentstyle=,
  stringstyle=\footnotesize\usefont{T1}{pcr}{m}{n}\selectfont,
  numberstyle=\footnotesize,
  tabsize=2,
  frame=lines,
  upquote=true,
  xleftmargin=10pt
}

\usepackage{xparse} 
\usepackage{stmaryrd} 

\NewDocumentCommand{\den}{oom}{%
  $
   \mbox{\bfseries#3}
   \IfValueT{#1}{_{#1}}
   \IfValueT{#2}{^{#2}}$%
}

\usepackage{cite}

\begin{document}
%

\bstctlcite{IEEEexample:BSTcontrol}

\title{AutoBotCatcher: Blockchain-based P2P Botnet Detection for the Internet of Things}


\author{
    \IEEEauthorblockN{Gokhan Sagirlar, Barbara Carminati, Elena Ferrari}
	\IEEEauthorblockA{University Of Insubria, Italy
    \{gsagirlar, barbara.carminati, elena.ferrari\}@uninsubria.it}
}



%


\maketitle
\thispagestyle{plain}
\pagestyle{plain}

\begin{abstract}
In general, a {\em botnet} is a collection of compromised internet computers, controlled by attackers for malicious purposes. To increase attacks' success chance and resilience  against defence mechanisms, modern botnets  have often a decentralized P2P structure. Here, IoT devices are playing a critical role, becoming one of the major tools for malicious parties to perform attacks. Notable examples are  DDoS attacks on Krebs on Security\footnote{krebsonsecurity.com/2016/09/krebsonsecurity-hit-with-record-ddos} and DYN\footnote{dyn.com/blog/dyn-analysis-summary-of-friday-october-21-attack}, which have been performed by IoT devices part of botnets.

We take a first step towards detecting P2P botnets in IoT, by proposing {\em AutoBotCatcher}, whose design is driven by the consideration that  bots of the same botnet frequently communicate with each other and form communities. 
As such,  the purpose of AutoBotCatcher is to dynamically analyze communities of IoT devices, formed according to their network traffic flows, to detect botnets.
AutoBotCatcher exploits a permissioned Byzantine Fault Tolerant (BFT) blockchain, as a state transition machine that allows collaboration of a set of pre-identified parties without trust, in order to perform collaborative and dynamic botnet detection by collecting and auditing IoT devices' network traffic flows as blockchain transactions.

In this paper, we focus on the design of the AutoBotCatcher by first defining the blockchain structure underlying  AutoBotCatcher, then discussing its  components.
\end{abstract}



%
\IEEEpeerreviewmaketitle

\section{Introduction}\label{sec:introduction}
IoT technology has been growing chaotically on many environments, such as homes and factories \cite{ourPaper}, connecting exceptionally large number of devices, expected to increase up to 130 billion devices in 2030.\footnote{\url{cdn.ihs.com/www/pdf/IoT_ebook.pdf}} 
Yet, this increasing popularity has made IoT devices a powerful amplifying platform for cyberattacks \cite{ddosInTheIoT}.
In fact, IoT devices are often simple products that due to the limits of  available constrained-resources  do not take security as primary goal.
As such, they represent a  rather easy target to attackers and the weakest link in the security chain of modern computer networks \cite{ddosInTheIoT}.
As proof of this, a recent study from HP found that more than 70\% of IoT devices do not have passwords with sufficient complexity and use unencrypted network services, resulting in being easy targets for  attackers.\footnote{\url{go.saas.hpe.com/l/28912/2015-07-21/32bhy3/28912/69168/IoT_Report.pdf}}

In such a vulnerable environment,  attackers can easily gain access to insecure IoT devices, and inject malicious softwares, \textit{malware}, to control them or to steal confidential information \cite{ourPaper2}. 
Today, one of the most relevant threat posed by malware in IoT  is represented by malicious \textit{botnets}.\footnote{For the rest of paper, we use the term botnet to refer malicious botnets.}
A botnet is a collection of compromised internet computers being controlled remotely by attackers for malicious and illegal purposes \cite{botnetCommunicationPatternsSurvey}.
For example, some recent Distributed Denial of Services (DDoS) attacks on {\em Krebs on Security} and {\em DYN} were due to  malware named Mirai \cite{mirai}, that uses IoT devices as botnets to generate extensive amount of network traffic, more than 1 Tbps. Additionally, such botnets have been commoditized by malicious parties, known as \textit{booters} \cite{stressTestingBooters},  that offers  DDoS as a service.
Booters exploit  compromised IoT devices to send attack packets to a target victim, in order to interrupt its service or shut it down.
Given that, botnets capable of using tens of thousands of IoT devices pose huge threats to online services' security and privacy.

\textbf{Botnets.} Let us examine in more details the main elements of botnets.
Briefly, a typical botnet consists of \cite{botnetCommunicationPatternsSurvey}: i) several \textit{bots}, that is,  infected machines running the bot executable; ii) a \textit{Command and Control (C\&C) server}, able to control every bot; and iii) a \textit{botmaster}, which is the malicious party controlling the botnet via the C\&C server.
Early botnets followed a centralized architecture, where the botmaster manages bots via the central C\&C server. To increase resilience of their attacks against defence mechanisms, more recent botnet architectures evolved into decentralized P2P architectures.
Today, decentralized P2P botnet topologies are able to utilize regular bots as C\&C servers   \cite{botnetCommunicationPatternsSurvey},  thus eliminating the single point of failure problem.
On the other hand, this makes P2P botnets harder to being detected and stopped, as botmasters are able to send attack commands through various channels.

\textbf{AutoBotCatcher}.  The design of AutoBotCatcher is driven by the consideration that  bots of a same botnet frequently communicate with each other and form communities \cite{friendsOfAnEnemyMutualContacts,peerhunter}. As such,  the purpose of AutoBotCatcher is to dynamically analyze communities of IoT devices, formed according to their network traffic flow (see problem statement in Section \ref{sec:system}), to detect botnets.
Specifically, it is a blockchain-based P2P botnet detection mechanism for IoT that makes use of two main actors, namely: \textit{agents} and \textit{block generators} (see  Section \ref{sec:system} for more details).
Where, agents are entitled to monitor  IoT network traffic flows in their subnets, and send  collected traffic information as blockchain transactions. 
In contrast,  by using collected  network traffic flows,  block generators (i.e., trusted big entities in IoT domain) aim at modeling  \textit{mutual contact} information of IoT devices (i.e., connections between IoT devices) and  generating  mutual contacts graph. 
This graph is then exploited to detect  communities (see Section \ref{sec:mutualContacts}).

In particular,  AutoBotCatcher uses Louvain method \cite{louvainAlgorithm} to perform community detection on mutual contacts graphs.
Since mutual contact information of IoT devices evolves over time, new snapshots of the mutual contacts graph are periodically generated. 
To this end, AutoBotCatcher exploits states of a BFT blockchain in order to store snapshots of the mutual contacts graph (see Section \ref{sec:paradigm}).
AutoBotCatcher's BFT blockchain is a permissioned blockchain, where a set of pre-identified block generators generate blocks and participate in the consensus process (see Section \ref{sec:blockchain}). Thus, network data stored on the blockchain is only accessible to block generators.

\textbf{Why blockchain?} 
To enable multiple parties to collaborate for botnet detection, we chose to use blockchain  rather than a centralized system given the benefits  blockchain might bring. 
Thanks to its distributed consensus protocol\footnote{With the assumption that more than two-thirds of the block generators are honest.}, blockchain platform  does not require a central trusted party to validate the correct execution of  the collaborative process (aka botnet detection), and ensure  transparency on  collected snapshots of communities of IoT devices overcoming the possible lack of trust among parties involved in the botnet detection (see Section \ref{sec:blockchain}). 
Moreover, as a state transition machine, blockchain lets us model the whole botnet detection process as a set of shared application states (aka states of parties collaborating in the  botnet detection).
This allows AutoBotCatcher to perform dynamic and collaborative botnet detection on large number of IoT devices.

\textbf{Contributions.} The main contributions described in this paper can be summarized as follows:\\
$\star$ a first blockchain-based botnet detection architecture for IoT;\\
$\star$ dynamic and collaborative approach for botnet detection and prevention; \\ 
$\star$ dynamic community detection with the help of blockchain technology.

\textbf{Outline.} The remainder of this paper is organized as follows.
In Section \ref{sec:system}, we present the considered  problem statement and the main entities involved in AutoBotCatcher.
We provide background information on blockchain technology, mutual contacts graph, and community detection approaches  in Section \ref{sec:background}.
In Section  \ref{sec:paradigm}, we introduce the blockchain paradigm defined for AutoBotCatcher.
We detail the design of AutoBotCatcher in Section \ref{sec:architecture}.
Section \ref{sec:related} discusses related work, whereas Section \ref{sec:conclusions} concludes the paper.

\section{System Model}\label{sec:system}
In this section, we introduce the  problem statement and the main entities of  AutoBotCatcher.

\textbf{Problem Statement.}
We assume to have a network of IoT devices, gateways, and some external hosts that communicate with IoT devices (such as device vendors' servers, cloud services).
IoT devices are connected to the internet, they sense and process data, and communicate with other IoT devices or external hosts. 
Botmasters compromise IoT devices and make them part of their botnets for malicious purposes, such as performing DDoS attacks.
On the other hand, gateways, such as Dell Edge Gateways\footnote{dell.com/us/business/p/edge-gateway}, are located in network boundaries and monitor the internet traffic to/from  IoT devices within their networks, referred to as their \textit{subnet}.
We assume that gateways are secure and trusted devices, 
as such, they behave as expected, and cannot be compromised by botmasters.

AutoBotCatcher targets P2P botnets, where all bots potentially can be utilized as C\&C servers by botmasters, and  performs community analysis on network traffic flows of IoT devices to detect botnet communities.
We assume that a botnet community is a group of compromised IoT devices
that frequently communicate with each other and with the same set of botmasters.
AutoBotCatcher relies on \textit{mutual contacts} information  of IoT devices, which refers to shared connections between a pair of IoT devices and/or other hosts.
For example, let us assume \textit{Host A}\footnote{Here, \textit{Host} refers to both IoT devices and other hosts in the network.} is connected to \textit{Host C}; given that, if \textit{Host B} is also connected to \textit{Host C} then \textit{Host A} and \textit{Host B} share a mutual contact, that is, \textit{Host C}.
As discussed in \cite{friendsOfAnEnemyMutualContacts} and \cite{peerhunter}, bots of a P2P botnet communicate with at least one mutual contact with very high probability, therefore mutual contacts can be exploited for botnet detection \cite{friendsOfAnEnemyMutualContacts}.
Given that, AutoBotCatcher exploits mutual contact information of IoT devices in performing botnet community analysis (see Section \ref{sec:architecture}).

Our assumptions on the threat model are as follows: IoT devices can become part of a botnet anytime; new types of botnets may emerge in the network; botmasters encrypt C\&C channels, and therefore DPI techniques are not suitable;  botnets tend to hide their operations and botmasters try to stay as stealthy as possible \cite{botnetCommunicationPatternsSurvey}, where botnets are able to manipulate characteristics of bot traffic, and thus they are able to make network flow traffic signature based defense approaches ineffective; botnets are in their waiting stage, where bots are joined to the C\&C network and wait for commands from the botmaster, thus their malicious activity may not be easily observable.  
The goal of AutoBotCatcher is to dynamically identify IoT devices and other hosts in the network that are part of botnets.

\textbf{Entities.}
AutoBotCatcher consists of two main entity types:

-- \textit{Agents:} They are typically gateway devices that are deployed in the network boundaries.
AutoBotCatcher's  agents monitor network traffic flows of IoT devices in their subnet and take actions, such as: generating network-data transactions (NTs) (see Definition \ref{def:NT}) and forcing infected devices to shut down. 
In AutoBotCatcher agents are considered trusted and honest. 

-- \textit{Block Generators:} This role is played by  big entities in IoT, such as device vendors, internet service providers (ISPs), security/privacy regulators.   
We assume that such big entities devote enough computing and network resources to take block generator role in P2P botnet detection process (see Section \ref{sec:architecture}).
For effective botnet detection and prevention, collaboration of different device vendors and ISPs is very important, as recent IoT botnets, such as Mirai and Hajime, infected IoT products from various vendors.\footnote{symantec.com/connect/blogs/hajime-worm-battles-mirai-control-internet-things} 
In fact, malware behind those botnets are adaptable to the various device architectures, such as ARM and Intel, and to different products, and they were effective in all around the world. 
Therefore, block generators collaborate to achieve large-scale defense and protection from botnet threat without trusting each other with the help of a permissioned BFT blockchain.
We assume that block generators have enough computing, storage and network resources available to devote to P2P botnet detection process operations using blockchain (see Section \ref{sec:architecture}).

\section{Background}\label{sec:background}
In this section, we provide  background information needed to understand the rest of the paper. To this end, first, we explain the mutual contacts graph concept; then, we provide an overview of community detection approaches; and, finally, we briefly describe blockchain technology.

\subsection{Mutual Contacts Graph}\label{sec:mutualContacts}
In AutoBotCatcher, mutual contacts graphs are exploited as a graph based data representation of  the mutual contacts of IoT devices and other hosts in the network. 
We denote a mutual contacts graph as $G = (V, E)$, where IP addresses of IoT devices and other hosts are \textit{vertices} ($V$). Vertices share an \textit{edge} ($E$), if they have at least one mutual contact.
Edges are bidirectional and weighted, where number of mutual contacts between vertices is the weight of the edge between them.
As such, by referring to the example of mutual contacts given in   Section \ref{sec:system} and assuming that there are only three hosts in the network, \textit{Hosts A, B, C} are vertices in the mutual contacts graph, where \textit{Host A} and \textit{Host B} share an edge with weight 1, as they share the mutual contact \textit{Host C}.
In AutoBotCatcher, the whole topology of the mutual contacts graph is represented by a 2 dimensional weighted adjacency matrix, referred to as \textit{mutual contacts matrix (MCM)}, whose element $MCM_{ij}$ indicates the number of mutual contacts between vertices $i$ and $j$.

\subsection{Community Detection}\label{sec:communityDetection}
Bots of the same botnet use similar C\&C channels and share the same messages \cite{peercleanUnveilingP2PbotnetsThroughDynamicGroupBehaviorAnalysis,peerhunter}, as such they are much likely to share many mutual contacts \cite{friendsOfAnEnemyMutualContacts} than legitimate P2P hosts. 
Therefore, P2P bots show community behaviours and form community structures that can be useful for botnet detection.
Given that, AutoBotCatcher performs \textit{community analysis} on mutual contacts graphs.

\textbf{Community detection methodology.} 
In AutoBotCatcher, accurate and fast detection of communities in the mutual contacts graph is  of great importance for proper botnet detection. 
Literature offers several community detection approaches for systems modeled as graphs (see \cite{communityDetectionInGraphs} for more details).
The main objective of these methods is to  find good partitions on the graphs as possible  communities in the network.
How to measure goodness of a partition is an important  concern in designing community detection methods. In general, a  \textit{quality function} is used as  a quantitative criterion that assigns a number to each partition of a graph to rank partitions based on their score \cite{communityDetectionInGraphs}.
Notably, \textit{modularity} is the most popular quality function, where achieving high modularity translates to a better partition of the graph, thus better community structure (for further discussion on the methodology please refer to \cite{findingEvaluatingCommunityStructureInNetworks}).
Given that, AutoBotCatcher exploits a modularity-based \textit{Louvain} method \cite{louvainAlgorithm}, that is, an hierarchical greedy algorithm trying to improve modularity for community detection.\footnote{Despite initially being designed for unweighted graphs, it can be easily adapted to weighted ones.} 
Louvain method has many advantages that makes it a good choice for AutoBotCatcher, such as: it is faster and achieves higher modularity than other methods; and it is able to process large networks in a short time.\footnote{It took 12 minutes for a network containing 39 million vertices and 783 million edges \cite{louvainAlgorithm} with AMD dual opteron 2.2k, 24 GB of RAM.}

\subsection{Blockchain}\label{sec:blockchain}
Blockchain relies on the concept of a distributed ledger maintained by a peer-to-peer network \cite{blockchainSurvey}. 
Novelty of the blockchain technology lies in its ability to achieve coordination and verification of individual activities carried out by different parties without a centralized authority or trusted third party, that allows decentralization of application execution with concerted and autonomous operations \cite{hybrid-iot}.
In blockchain, \textit{transactions} transfer information (i.e., data packets) between peers. They have a unique identifier (transaction-id), input data, and are bundled into data chunks, referred to as \textit{blocks}. 
Block generator peers of the blockchain broadcast blocks by exploiting public-key cryptography.
Blocks are recorded in the blockchain with an exact order. 
Briefly, a block contains: a set of transactions; a timestamp; a reference to the preceding block that identifies the block's place in the blockchain; an authenticated data structure (e.g., a Merkle tree) to ensure block integrity.\footnote{Block structure varies in different blockchain protocols, here we list the most common elements.}

As a state transition machine, different participants in the blockchain have to achieve consensus on the latest state of the ledger in order to achieve coordination on the processes that they perform on the ledger. There are different methodologies to achieve consensus in the blockchain.
For example, some blockchains use  \textit{Byzantine Fault Tolerant (BFT)} methodologies that depend on state replication between block generators;
other use \textit{Proof of Work (PoW)}, where block generators have to use their hardware resources and energy to generate a block by solving a cryptographic puzzle; while others use \textit{Proof of Stake (PoS)}, where block generator selection depends on part of peers' wealth that they have voted as their stake in the block generator selection process.
Notably, BFT blockchains can maintain a relative high throughput.
For example, BFT Tendermint consensus protocol  \cite{tendermint}  is able to process thousands of transactions per second.
Generally, BFT blockchains can scale to dozens or few hundreds of block generators.

Blockchains can be classified into two groups as {\it public} and {\it permissioned} according to their way of regulating peers' participation in blockchain operations. 
Particularly, in public blockchains, any peer can read and write to the blockchain, meaning that anyone can participate in the consensus process.
Whereas, in permissioned blockchains, only a set of previously identified peers can write to the blockchain and participate in the consensus and access to the stored transactions. 

In AutoBotCatcher, we exploit a BFT blockchain in permissioned setting for processing network traffic flow data for botnet community detection.
Permissioned BFT blockchain is adequate for AutoBotCatcher setting, since its design is based on assigning the block generator role to relatively small set of actors such as device vendors, internet service providers etc. 
Moreover, AutoBotCatcher exploits states of a BFT blockchain in performing dynamic botnet community detection process, as discussed in the following section.
Finally, in permissioned blockchain setting stored network data kept confidential from unknown parties as only a set of pre-identified block generators can access to network data stored in the blockchain.

\section{The Blockchain Paradigm}\label{sec:paradigm}
We devote this section to explain how blockchain is used in AutoBotCatcher. 

\textbf{Transactions.} To detect botnets, AutoBotCatcher employs community detection analysis on the mutual contacts graph. In particular, to generate mutual contacts graphs, AutoBotCatcher needs  to collect and analyze meta-data information about network traffic flow of IoT devices (i.e., IP addresses).
To this end, we exploit a permissioned blockchain as a shared data store to audit meta-data, which are thus modelled as blockchain transactions sent by agents to the blockchain's \textit{transaction pool}, a shared data store hosted by all block generators, that holds unprocessed transactions. Particularly, each agent is connected to one block generator to send transactions, where block generator that receives a new transaction disseminates new transactions to other block generators.

Meta-data are encoded into \textit{network-data transaction (NT)}, formally defined as follows: 

\begin{definition} \textbf{Network-data Transaction (NT).}\label{def:NT}
Let $NetFlow$ be the network traffic flow  defined as $NetFlow=(IP_{src},IP_{dest})$, where  $IP_{src}$ and $IP_{dest}$ are the source's and destination's IP addresses, respectively. 
Let $Device_{addr}$ be the unique public key of the agent that sends the transaction, 
and let $Tx-Pool_{addr}$ be the public key of the transaction pool that agent is connected to send the transaction,
A network-data transaction is a tuple: $NT = \langle Device_{addr}, IP_{src}, IP_{dest}, Tx-Pool_{addr} \rangle$
\end{definition}

\textbf{Blocks.} Transactions are bundled into  blocks that are generated by block generators.\footnote{Number of transactions in  a block is regulated according to the block size and transaction size settings of the blockchain protocol.}  In what follows, we symbolize a block as $b_{m}$, where $m$ represents the block number. Given that, the block structure is  as follows:
\begin{align}\label{eq:block}
b_{m} = \{ NT_{x}, \dots, NT_{y}\}
\end{align}

Where, $NT$ represents a network-data transaction, and $x$ and $y$ are the transaction number.
During the execution of  AutoBotCatcher, one of the block generators is elected as a leader to generate a block. 
Block generation interval, symbolized as $\tau$, represents the amount of time between consecutive block generations.
Upon generation of a block, at least two-thirds of block generators should send acknowledgements to the block generator regarding their approval on the block.\footnote{If the block generator does not get approval from at least two-thirds of block generators, that block will not be added to the blockchain, and a new block generator will be selected to generate a block.}
Once acknowledgements have been obtained and $\tau$ is expired, block generator of the next block generates the new block.

\textbf{Rounds.} In AutoBotCatcher, P2P botnet analysis is periodically performed upon generation of a set of blocks.
We refer to such periods  as  \textit{rounds}.
Each round takes certain amount of time, symbolized as $\Delta$.
Value of  $\Delta$ depends on both the amount of time needed for block generators to perform botnet detection operations on the blocks (e.g., mutual contacts graph extraction, botnet community detection, etc.),  network quality (e.g., network connection delays), and blockchain protocol (e.g., transaction throughput etc.).
A round is symbolized as $\Gamma_{\Delta_{t}}$, where $\Delta_{t}$ is a timestamp specifying the starting time of the round.
When the round $\Gamma_{\Delta_{t}}$ ends, the next round $\Gamma_{\Delta_{t+1}}$ starts. 
In round $\Gamma_{\Delta_{t}}$, botnet detection is performed on NTs sent during the previous round $\Gamma_{\Delta_{t-1}}$.
More precisely, NTs sent from timestamp $\Delta_{t-1}$ to timestamp $\Delta_{t}$ (which corresponds to execution time of round $\Gamma_{\Delta_{t-1}}$) are processed in round $\Gamma_{\Delta_{t}}$.

\textbf{State.} 
Fundamentally, blockchain consists of set of shared states and performs state transition.
In general, the \textit{state} notion can be used to represent financial balances of users (e.g., Bitcoin), or, in a broader setting,  it can represent anything that can be modeled as result of the computer programs (i.e., on Ethereum blockchain \cite{ethereumYellowpaper}). 
In our setting, the state is a mapping between IoT devices' and other hosts' IP addresses (that are subject to botnet detection) and communities (each marked as botnet community or benign community).
In AutoBotCatcher, similar to the  Ethereum protocol \cite{ethereumYellowpaper}, the state is not stored on the blockchain, rather it is maintained on an efficient Merkle tree implementation,\footnote{Exploiting Merkle trees in our blockchain setting requires a simple state database backend.} such as Patricia Merkle Trees.\footnote{github.com/ethereum/wiki/wiki/Patricia-Tree}
Merkle trees are hash based data structures, where each node is the hash of its children or hash of the data, if the node is a leaf.
Main benefits of using Merkle trees to store states are: they are  immutable data structures, thus we are able to secure entire system states with cryptographic dependences; and, they allow the blockchain protocol to trivially revert to any old state by simply altering the root hash \cite{ethereumYellowpaper}.

In AutoBotCatcher, the state of the blockchain on timestamp $\Delta_{t}$, symbolized as $\sigma_{\Delta_{t}}$, consists of: the snapshot of the latest version of the mutual contacts graph $G_{\Delta_{t}}$, that is generated from the network-data transactions;
the set of all communities and  IP addresses of hosts associated with that communities, extracted from the mutual contacts graph by block validators,  symbolized as $CommSet_{\Delta_{t}}$;
and, all blocks of the blockchain.
Given that, the state of the blockchain on timestamp $\Delta_{t}$ is represented as follows:

\begin{align}\label{eq:state}
\sigma_{\Delta_{t}} = \left( G_{\Delta_{t}}, CommSet_{\Delta_{t}}, [B_{0},\dots,B_{m}] \right) 
\end{align}

\begin{figure} [h] 
	\centering
	\includegraphics[width=86mm]{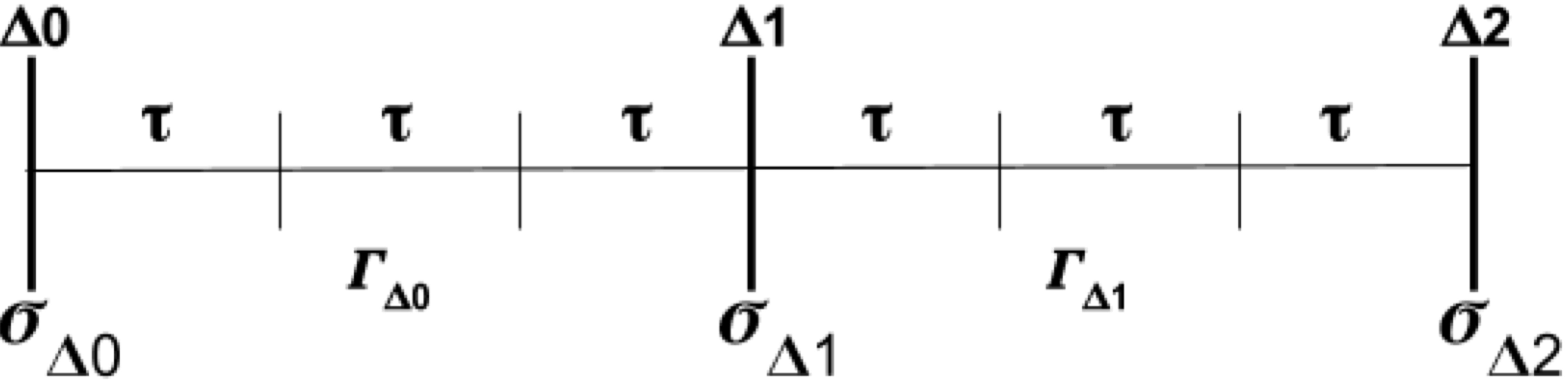}
	\caption{Round and state relation} \label{fig:round}
\end{figure}

In Figure \ref{fig:round}, we present an example  of rounds and states, where each round includes generation of three blocks.

\textbf{State transitions.} In  blockchain, state transition refers to achieving a new valid state after executing a set of transactions on the previous state. 
AutoBotCatcher uses the community set and snapshot of the mutual contacts graph of the previous blockchain state to  dynamically perform botnet community detection.
Therefore, in AutoBotCatcher, a state transition occurs upon finishing a round of botnet detection operations (see Section \ref{sec:architecture} for more details).
Given that, for every round $\Gamma$ defined above, one state change occurs upon execution of the set of blocks. 
Particularly, upon execution of round $\Gamma_{\Delta_{t}}$, the state changes from $\sigma_{\Delta_{t}}$ to $\sigma_{\Delta_{t+1}}$ which includes a new set of processed blocks.
We define the state transition as a function of rounds, which takes previous state and new blocks as input, as follows:

\begin{align}
\sigma_{\Delta_{t+1}} = \Gamma\left(\sigma_{\Delta_{t}}, [B_{m}, \dots, B_{m+z}]\right)
\end{align}

Where $\sigma_{\Delta_{t}}$ is the previous state (see Equation \ref{eq:state}), and $[B_{m}, \dots, B_{m+z}]$ is the set of new blocks, where $m$ and $z$  represent block number (see Equation \ref{eq:block}).

\textbf{Consensus.} 
Blockchain protocol adopted by AutoBotCatcher uses Byzantine Fault Tolerant (BFT) methodology to achieve consensus in a permissioned setting.
In BFT blockchain, in order to achieve consensus, at least two-thirds of the pre-identified block generators have to agree on the latest state proposed by one of the block generators. 
In our setting, this translates to  agree on: the same set of blocks processed; same mutual contacts graph; and same community mapping of IoT devices.

\section{System Architecture}\label{sec:architecture}
\begin{figure*} [h] 
	\centering
	\includegraphics[width=\textwidth]{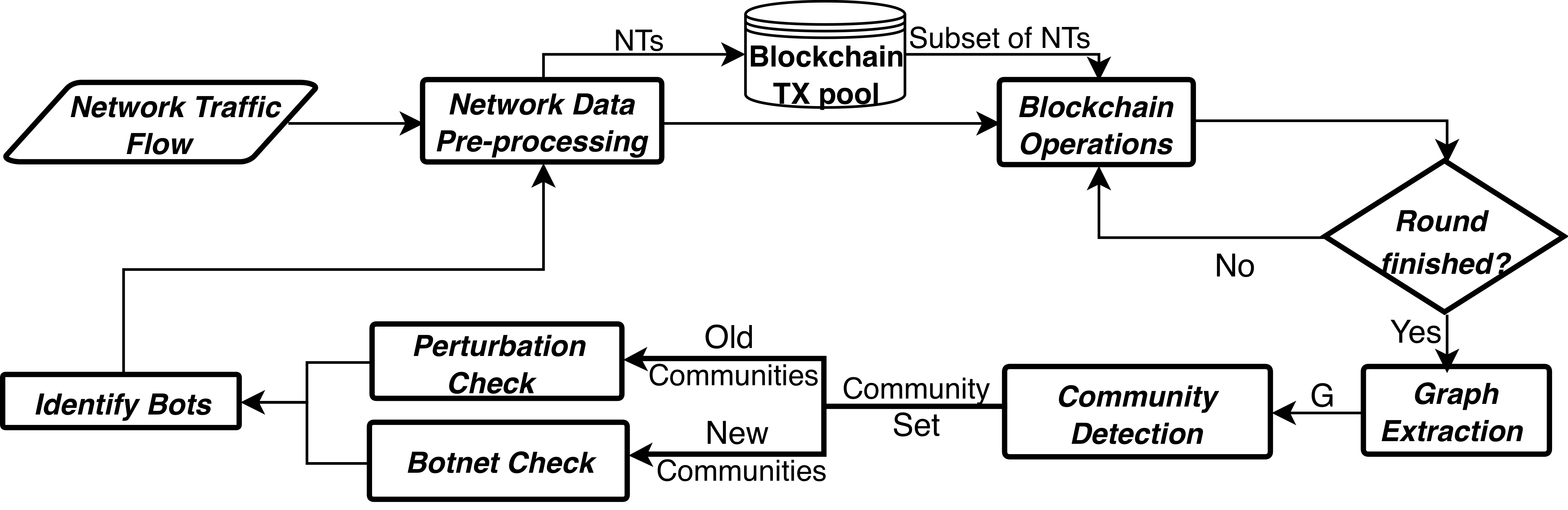}
    \caption{AutoBotCatcher system flow} \label{fig:flow}
\end{figure*}

AutoBotCatcher exploits a permissioned BFT blockchain as  a backend module to perform dynamic and collaborative botnet detection on large scale networks. 
An overview of the execution  flow of  AutoBotCatcher is presented in Figure \ref{fig:flow} and discussed in what follows.

\textbf{Network data pre-processing.}
This task  is executed by agents  for monitoring network traffic flow of IoT devices and taking actions according to that. 
In performing such operations, agents maintain two types of IP address lists, namely \textit{blacklist} and \textit{whitelist}.
Blacklists contain IP addresses that have been previously detected as  part of a botnet. 
On the other hand, whitelists contain predefined and trusted IP addresses,\footnote{We assume that hosts corresponding to IP addresses in whitelists are secure, and do not pose any threat to IoT devices.} such as IP addresses of the vendor's servers, for all IoT devices in their subnet. 

More precisely,  an agent constantly sniffs network traffic flows of IoT devices in its subnet; the agent does not take any action for network traffic of an IoT device with a whitelisted IP address; for all other network traffic flows, the agent forms network flow transactions (NTs) (as given in Definition \ref{def:NT}), if a network flow is with a blacklisted IP address, then agent quarantines that IoT device either by forcing it to shut down or by cutting all of its network connections.

\textbf{Blockchain operations.} 
This task is executed by block generators for the generation and relay of blocks that include NTs.
To perform these operations in a round $\Gamma_{\Delta_{t}}$, first, one block generator is  selected to generate one or more blocks.\footnote{For the sake of brevity, we do not detail the election process, but any leader election process performed in a typical distributed system is suitable for our setting.}
That block generator takes the subset of NTs from the blockchain transaction pool, it forms a block and broadcasts the block to all block generators.
We recall that each block has to receive approval from at least two-thirds of the block generators in order to be valid, and to achieve consensus on its validity in a BFT blockchain setting.

\textbf{Graph extraction.}
This task is  executed by block generators, aiming at elaborating the  network traffic flow data, that is,  the NTs processed in a round $\Gamma$, to generate/update the mutual contacts graph $G$. Particularly in round $\Gamma_{\Delta_{t}}$, block generators create the mutual contacts matrix $MCM_{\Delta_{t+1}}$,  representing $G_{\Delta_{t+1}}$ that results from the state $\sigma_{\Delta_{t+1}}$, by updating $MCM_{\Delta_{t}}$, i.e.,  $G_{\Delta_{t}}$ of state $\sigma_{\Delta_{t}}$, with the NTs in blocks sent from $\Delta_{t-1}$ to $\Delta_{t}$. 

Operations to transit from $MCM_{\Delta_{t}}$ to $MCM_{\Delta_{t+1}}$ are: \textit{updating weights of edges} between vertices that communicated in round $\Gamma_{\Delta_{t}}$; \textit{adding vertices and edges},
if new IoT devices connect to a device network, or a new host IP address communicates with an IoT device; \textit{removing vertices and edges}, if some IoT devices or hosts do not exist anymore.

\textbf{Community detection.}
This task, executed by block generators,  performs  dynamic community discovery (DCD) on the mutual contacts graph $G$.

AutoBotCatcher performs dynamic community detection with Louvain method on snapshots of the mutual contacts graphs from consecutive states of the permissioned BFT blockchain.
Yet, as presented in \cite{staticCommunityDetectionAlgorithmsForEvolvingNetworks}, even small changes in consecutive network snapshots may cause Louvain method to generate two alike community structures, which would eventually cause AutoBotCatcher to lose track of the communities and thus degrade its execution.
Therefore, to have a more stable community structure between timestamps, AutoBotCatcher initializes the Louvain algorithm for community detection at timestamp $\Delta_{t+1}$ with the communities found in $\Delta_{t}$ as proposed by Aynaud \textit{et al.} in \cite{staticCommunityDetectionAlgorithmsForEvolvingNetworks}. 
More precisely, in round $\Gamma_{\Delta_{t}}$, AutoBotCatcher performs community detection on newly extracted mutual contacts graph $G_{\Delta_{t+1}}$, by feeding community set of IoT devices, $CommSet_{\Delta_{t}}$ from the last state $\delta_{\Delta_{t}}$ to Louvain method.

\textbf{Perturbation check.}
This task is performed by  block generators to check updates (i.e., IP address additions and removals) on communities that already exist in the previous timestamp.\footnote{If a community changes more than a threshold $\phi$, calculated as the ratio of total changes to number of edges and vertices, that community will be considered as a new community, and next task will be executed on it.}
Particularly, for botnet communities, IP addresses of new bots are inserted to the list called \textit{additions to blacklist}, whereas  IP addresses of bots that are no longer part of a botnet are inserted to the list called \textit{removals from blacklist}.
Upon block generators achieve consensus on the new state, block generators share those lists with agents (see task bot identifier).

\textbf{Botnet check.}
Executed by block generators, this task is responsible for classifying new communities as either botnet or benign communities.
Botnet detection methodology used by AutoBotCatcher is based on two observations: 1) bots  connect with each other for command exchange, thus they have more mutual contacts than benign communities \cite{peerhunter}; 2) botmasters or attack targets communicate with many nodes, referred to as pivotal nodes \cite{botnetDetectionAnomalyCommunityDetection}, so that they have very high number of mutual contacts.  
Therefore, according to the first observation, AutoBotCatcher calculates the average number of mutual contacts per IP address for all communities. If the result is higher than a pre-defined mutual contacts threshold $\theta$, that community is marked as \textit{candidate botnet} community.
Secondly, AutoBotCatcher checks for pivotal nodes in candidate botnet communities. Particularly, for all IP addresses in a candidate botnet community, AutoBotCatcher sums their rows in the mutual contacts matrix (MCM). If one or more pivotal nodes exists in the candidate botnet community, it is marked as botnet community by the block generator.
Upon block generators achieve consensus on the new state, block generators share IP addresses in botnet communities with next task for bot blacklisting.

\textbf{Identify bots.}
The last task is responsible for updating the bot blacklists of agents, and it is executed by block generators after a state change.
It updates the blacklists of all IoT devices with the help of smart contract transactions for addition and deletion of IP addresses on agent's local blacklists.\footnote{As it is not the main focus of our work, we do not detail how this mechanism works, such as transaction structure, which is left as future work.}

\section{Related Work}\label{sec:related}
In recent years, vast amounts of work has been devoted to P2P botnet detection. In general, botnet detection methodologies can be categorized into two groups: host-based and network-based approaches.  Host-based approaches require the  monitoring of all hosts, which is impractical for the IoT domain. Therefore, we focus on network-based approaches,  which  in turn  can be classified into two groups:

\textit{-- Network traffic signature based approaches.} 
Literature offers many works that classify hosts based on their network traffic behaviour. In general, these approaches exploit supervised/unsupervised machine learning techniques to identify whether hosts are benign or malicious \cite{botminer,hostsTrading,detectingStealthyP2PbotnetsUsingStatisticalTrafficFingerprints,detectingP2PbotnetsNetworkBehavAnalysisML,peerrush,scalableSystemForStealthyP2PbotnetDetection,peershark,botdetector,detectingp2pBotnetSDN}.
However, in a dynamic network, botmasters can randomize botnet traffic by changing communication frequency, packet sizes, etc. 
As such, network traffic signatures learned by machine learning approaches may not be robust enough to identify bots \cite{peercleanUnveilingP2PbotnetsThroughDynamicGroupBehaviorAnalysis}, which would eventually make such approaches ineffective.

Moreover, some of the proposed approaches, such as \cite{botminer, botdetector}, rely on deep packet inspection techniques (DPI) to analyze network packet contents. However, these checks can be bypassed through encryption of C\&C channels. Given that, we conclude that  network traffic signature based approaches are not suitable for dynamic and evolving IoT environments.

\textit{-- Group and community behavior based approaches.} 
Some works  use group and community behaviour analysis for botnet detection \cite{friendsOfAnEnemyMutualContacts, peercleanUnveilingP2PbotnetsThroughDynamicGroupBehaviorAnalysis,peerhunter}.
As an example, similarly to us, \cite{friendsOfAnEnemyMutualContacts} exploits mutual contacts  extracted from network traffic flow of hosts, in order to identify bots in a P2P network. 
Whereas \cite{peercleanUnveilingP2PbotnetsThroughDynamicGroupBehaviorAnalysis} performs group level behavior analysis on network traffic flow with Support Vector Machine (SVM). 
However, these approaches are able  to detect only  previously known bot types.
Therefore, they are not suitable for IoT, where new botnets emerge frequently \cite{ddosInTheIoT}.
Differently, PeerHunter  \cite{peerhunter} exploits Louvain method to perform network flow level community behaviour analysis on mutual contacts graph, without relying on previously known bot types.
Yet, PeerHunter performs static botnet detection on the collected network traffic flow data, which is inadequate for a dynamic and evolving IoT environment that requires dynamic botnet detection.

AutoBotCatcher differs from the previously discussed approaches in several ways.  
First, unlike DPI based approaches, to perform botnet detection AutoBotCatcher requires to trace only high level meta-data about network flow traffic  (i.e., source and destination addresses), as such it is effective against encrypted C\&C channels.
Second, unlike  network traffic signature based approaches, even if botmasters randomize network traffic by changing packet sizes, communication frequency etc.,  botnet community structures  do not alter, since the same set of commands has to be shared with the same set of bots.
Third, unlike other group and community behaviour based approaches, AutoBotCatcher performs  dynamic community detection by exploiting blockchain, so it is able to detect emerging and unknown botnets, and to take preventive measures against them.

\section{Conclusions and Future Work}\label{sec:conclusions}
In this paper, we introduce a blockchain based dynamic P2P botnet detection and prevention mechanism for IoT, referred to as AutoBotCatcher, that performs community detection on network flows of IoT devices. 
In AutoBotCatcher,  IoT gateway devices become peers of a BFT blockchain, where device vendors and/or security regulators take the block generator role and participate in the consensus process.
Blockchain is exploited to perform dynamic network  based botnet community  detection on snapshots of the mutual contact graph of IoT devices.

Future work includes: implementing AutoBotCatcher by leveraging on BFT based Hyperledger blockchain\footnote{hyperledger.org} and exploiting its smart contracts to generate mutual contacts graphs and to manage state changes;
integrating secure multi-party computation platforms, such as Enigma \cite{enigma}, in order to protect privacy of the collaborating parties,
while ensuring correct botnet detection;
testing  AutoBotCatcher with several botnet examples; 
designing and implementing  a trust management module between agents and block generators, in order to make AutoBotCatcher resilient against insider threats. We also plan to extend the current botnet detection model with methods making use of statistical network traffic features.


\bibliographystyle{IEEEtran}
\bibliography{biblio}

\begin{thebibliography}{10}
\providecommand{\url}[1]{#1}
\csname url@samestyle\endcsname
\providecommand{\newblock}{\relax}
\providecommand{\bibinfo}[2]{#2}
\providecommand{\BIBentrySTDinterwordspacing}{\spaceskip=0pt\relax}
\providecommand{\BIBentryALTinterwordstretchfactor}{4}
\providecommand{\BIBentryALTinterwordspacing}{\spaceskip=\fontdimen2\font plus
\BIBentryALTinterwordstretchfactor\fontdimen3\font minus
  \fontdimen4\font\relax}
\providecommand{\BIBforeignlanguage}[2]{{%
\expandafter\ifx\csname l@#1\endcsname\relax
\typeout{** WARNING: IEEEtran.bst: No hyphenation pattern has been}%
\typeout{** loaded for the language `#1'. Using the pattern for}%
\typeout{** the default language instead.}%
\else
\language=\csname l@#1\endcsname
\fi
#2}}
\providecommand{\BIBdecl}{\relax}
\BIBdecl

\bibitem{ourPaper}
B.~Carminati \emph{et~al.}, ``Enhancing user control on personal data usage in
  internet of things ecosystems,'' in \emph{Services Computing (SCC), 2016 IEEE
  International Conference on}.\hskip 1em plus 0.5em minus 0.4em\relax IEEE,
  2016, pp. 291--298.

\bibitem{ddosInTheIoT}
C.~Kolias \emph{et~al.}, ``Ddos in the iot: Mirai and other botnets,''
  \emph{Computer}, vol.~50, no.~7, pp. 80--84, 2017.

\bibitem{ourPaper2}
G.~{Sagirlar} \emph{et~al.}, ``{Decentralizing Privacy Enforcement for Internet
  of Things Smart Objects},'' \emph{ArXiv e-prints}, Apr. 2018.

\bibitem{botnetCommunicationPatternsSurvey}
G.~Vormayr \emph{et~al.}, ``Botnet communication patterns,'' \emph{IEEE
  Communications Surveys \& Tutorials}, vol.~19, no.~4, pp. 2768--2796, 2017.

\bibitem{mirai}
M.~Antonakakis \emph{et~al.}, ``Understanding the mirai botnet,'' in
  \emph{USENIX Security Symposium}, 2017.

\bibitem{stressTestingBooters}
M.~Karami \emph{et~al.}, ``Stress testing the booters: Understanding and
  undermining the business of ddos services,'' in \emph{Proceedings of the 25th
  International Conference on World Wide Web}.\hskip 1em plus 0.5em minus
  0.4em\relax International World Wide Web Conferences Steering Committee,
  2016, pp. 1033--1043.

\bibitem{friendsOfAnEnemyMutualContacts}
B.~Coskun \emph{et~al.}, ``Friends of an enemy: identifying local members of
  peer-to-peer botnets using mutual contacts,'' in \emph{Proceedings of the
  26th Annual Computer Security Applications Conference}.\hskip 1em plus 0.5em
  minus 0.4em\relax ACM, 2010, pp. 131--140.

\bibitem{peerhunter}
D.~Zhuang \emph{et~al.}, ``Peerhunter: Detecting peer-to-peer botnets through
  community behavior analysis,'' in \emph{Dependable and Secure Computing, 2017
  IEEE Conference on}.\hskip 1em plus 0.5em minus 0.4em\relax IEEE, 2017, pp.
  493--500.

\bibitem{louvainAlgorithm}
V.~D. Blondel \emph{et~al.}, ``Fast unfolding of communities in large
  networks,'' \emph{Journal of statistical mechanics: theory and experiment},
  vol. 2008, no.~10, p. P10008, 2008.

\bibitem{peercleanUnveilingP2PbotnetsThroughDynamicGroupBehaviorAnalysis}
Q.~Yan \emph{et~al.}, ``Peerclean: Unveiling peer-to-peer botnets through
  dynamic group behavior analysis,'' in \emph{Computer Communications
  (INFOCOM), 2015 IEEE Conference on}.\hskip 1em plus 0.5em minus 0.4em\relax
  IEEE, 2015, pp. 316--324.

\bibitem{communityDetectionInGraphs}
S.~Fortunato, ``Community detection in graphs,'' \emph{Physics reports}, vol.
  486, no. 3-5, pp. 75--174, 2010.

\bibitem{findingEvaluatingCommunityStructureInNetworks}
M.~E. Newman \emph{et~al.}, ``Finding and evaluating community structure in
  networks,'' \emph{Physical review E}, vol.~69, no.~2, p. 026113, 2004.

\bibitem{blockchainSurvey}
F.~Tschorsch \emph{et~al.}, ``Bitcoin and beyond: A technical survey on
  decentralized digital currencies,'' \emph{IEEE Communications Surveys \&
  Tutorials}, vol.~18, no.~3, pp. 2084--2123, 2016.

\bibitem{hybrid-iot}
G.~{Sagirlar} \emph{et~al.}, ``{Hybrid-IoT: Hybrid Blockchain Architecture for
  Internet of Things - PoW Sub-blockchains},'' \emph{ArXiv e-prints}, Apr.
  2018.

\bibitem{tendermint}
E.~Buchman, ``Tendermint: Byzantine fault tolerance in the age of
  blockchains,'' Ph.D. dissertation, 2016.

\bibitem{ethereumYellowpaper}
G.~Wood, ``Ethereum: A secure decentralised generalised transaction ledger,''
  \emph{Ethereum Project Yellow Paper}, 2016.

\bibitem{staticCommunityDetectionAlgorithmsForEvolvingNetworks}
T.~Aynaud \emph{et~al.}, ``Static community detection algorithms for evolving
  networks,'' in \emph{Modeling and optimization in mobile, ad hoc and wireless
  networks (WiOpt), 2010 proceedings of the 8th international symposium
  on}.\hskip 1em plus 0.5em minus 0.4em\relax IEEE, 2010, pp. 513--519.

\bibitem{botnetDetectionAnomalyCommunityDetection}
J.~Wang \emph{et~al.}, ``Botnet detection based on anomaly and community
  detection,'' \emph{IEEE Transactions on Control of Network Systems}, vol.~4,
  no.~2, pp. 392--404, 2017.

\bibitem{botminer}
G.~Gu \emph{et~al.}, ``Botminer: Clustering analysis of network traffic for
  protocol-and structure-independent botnet detection.'' in \emph{USENIX
  security symposium}, vol.~5, no.~2, 2008, pp. 139--154.

\bibitem{hostsTrading}
T.-F. Yen \emph{et~al.}, ``Are your hosts trading or plotting? telling p2p
  file-sharing and bots apart,'' in \emph{Distributed Computing Systems
  (ICDCS), 2010 IEEE 30th International Conference on}.\hskip 1em plus 0.5em
  minus 0.4em\relax IEEE, 2010, pp. 241--252.

\bibitem{detectingStealthyP2PbotnetsUsingStatisticalTrafficFingerprints}
J.Zhang \emph{et~al.}, ``Detecting stealthy p2p botnets using statistical
  traffic fingerprints,'' in \emph{Dependable Systems \& Networks (DSN), 2011
  IEEE/IFIP 41st International Conference on}.\hskip 1em plus 0.5em minus
  0.4em\relax IEEE, 2011, pp. 121--132.

\bibitem{detectingP2PbotnetsNetworkBehavAnalysisML}
S.~Saad \emph{et~al.}, ``Detecting p2p botnets through network behavior
  analysis and machine learning,'' in \emph{Privacy, Security and Trust (PST),
  2011 Ninth Annual International Conference on}.\hskip 1em plus 0.5em minus
  0.4em\relax IEEE, 2011, pp. 174--180.

\bibitem{peerrush}
B.~Rahbarinia \emph{et~al.}, ``Peerrush: Mining for unwanted p2p traffic,'' in
  \emph{International Conference on Detection of Intrusions and Malware, and
  Vulnerability Assessment}.\hskip 1em plus 0.5em minus 0.4em\relax Springer,
  2013, pp. 62--82.

\bibitem{scalableSystemForStealthyP2PbotnetDetection}
J.~Zhang \emph{et~al.}, ``Building a scalable system for stealthy p2p-botnet
  detection,'' \emph{IEEE transactions on information forensics and security},
  vol.~9, no.~1, pp. 27--38, 2014.

\bibitem{peershark}
P.~Narang \emph{et~al.}, ``Peershark: detecting peer-to-peer botnets by
  tracking conversations,'' in \emph{Security and Privacy Workshops (SPW), 2014
  IEEE}.\hskip 1em plus 0.5em minus 0.4em\relax IEEE, 2014, pp. 108--115.

\bibitem{botdetector}
S.~Mizuno \emph{et~al.}, ``Botdetector: A robust and scalable approach toward
  detecting malware-infected devices,'' in \emph{Communications (ICC), 2017
  IEEE International Conference on}.\hskip 1em plus 0.5em minus 0.4em\relax
  IEEE, 2017, pp. 1--7.

\bibitem{detectingp2pBotnetSDN}
S.-C. Su \emph{et~al.}, ``Detecting p2p botnet in software defined networks,''
  \emph{Security and Communication Networks}, vol. 2018, 2018.

\bibitem{enigma}
G.~Zyskind \emph{et~al.}, ``Enigma: Decentralized computation platform with
  guaranteed privacy,'' \emph{arXiv preprint arXiv:1506.03471}, 2015.

\end{thebibliography}

\end{document}